# Bivariate Inverse Topp-Leone Model to Counter Heterogeneous Data

Shikhar Tyagi
Department of Mathematics, GITAM University, Bangalore, India.
Email: shikhar1093tyagi@gmail.com

**Abstract**

In probability and statistics, reliable modeling of bivariate continuous characteristics remains a real insurmountable consideration. During analysis of bivariate data, we have to deal with heterogeneity that is present in data. Therefore, for dealing with such a scenario, we investigate a novel technique based on a Farlie-Gumbel-Morgenstern (FGM) copula and the inverse Topp-Leone (ITL) model in this study. The idea is to use the oscillating functionalities of the FGM copula and the flexibility of the ITL model to propose a serious bivariate solution for the modeling of bivariate lifetime phenomena to counter the heterogeneity present in data. Both theory and practice are developed. In particular, we determine the main functions related to the model, like the cumulative model function, probability density function, conditional density function, and various useful dependence measures for bivariate modeling. The model parameters are estimated using the maximum likelihood method and Bayesian framework of Markov Chain Monte Carlo (MCMC) methodology. Following that, model comparison methods are used to compare models. To explain the findings and show that better models are recommended, the famous Drought and Burr data sets are used.

## 1 Introduction

Classical probability models are important throughout many domains of applied research, including reliability, economics, medical sciences, and other advanced disciplines. For assessing lifetime data, the gamma and exponential distributions are often used in probability distributions. In the literature, several extensions of the gamma and exponential distributions, as well as their mixtures, have been proposed and explored, and have been effectively used for modeling and understanding different lifespan phenomena (see Johnson *et al.* (1994); Sarhan and Kundu (2009); Sen *et al.* (2016)). The classical distributions have constraints when dealing with a large range of real-world data, which motivates the development of new flexible distribution families. Various methods for creating bivariate distributions from conventional univariate distributions have been demonstrated in recent times. Numerous distributions have been suggested for the study of bivariate lifetime data, which extend several prominent univariate distributions including exponential, Weibull, Pareto, gamma, and log-normal distributions. (see, for example, Gumbel (1960); Marshall and Olkin (1967); Sankaran and Nair (1993); Kundu and Gupta (2009); Sarhan *et al.* (2011)). The formation of bivariate distributions employing conditional and marginal distributions is a suitable strategy that has received a lot of attention in recent years. Several magnificent approaches for generating bivariate distributions through order statistics had recently been presented and researched, which contains both absolutely continuous and singular components and may be advantageous in circumstances when data ties exist. For some recent references, one can refer to Dolati *et al.* (2014), Mirhosseini *et al.* (2015), and Kundu and Gupta (2017). Copula models have lately been used to describe the dependency between random variables, in addition to current methodologies. A copula is a function that connects the marginals to the joint distribution and has been widely utilized in finance, biology, engineering, hydrology, and geophysics to explain dependency among random variables. On the unit interval $[0,1]$, a copula is a multivariate distribution function with uniform one-dimensional margins. In this paper, we restrict our study to a bivariate copula. A formal definition of the bivariate copula is as follows:

A function $C:[0,1]\times[0,1]\to[0,1]$ is a bivariate copula if it satisfies the following properties:

i. For every $u,v \in [0,1]$

$$C(u,0) = 0 = C(0,v)$$

and

$$C(u,1) = 1 \text{ and } C(1,v) = v$$

ii. For every $u_1, u_2, v_1, v_2 \in [0,1]$ such that $u_1 \leq u_2$ and $v_1 \leq v_2$

$$C(u_2, v_2) - C(u_2, v_1) - C(u_1, v_2) + C(u_1, v_1) \geq 0.$$

Let $X$ and $Y$ be random variables with joint distribution function $F$, and marginals $F_1$ and $F_2$, respectively, then {\cite{sklar1959fonctions}} says that there exists a copula function $C$ which connects marginals to the joint distribution via the relation $F(x,y) = P(X \leq x, Y \leq y) = C(F_1(x), F_2(y))$. If $X$ and $Y$ are continuous, then the copula $C$ is unique; otherwise, it is uniquely determined on $\text{Range}(F_1) \times \text{Range}(F_2)$. The associated joint density is $f(x,y) = c(F_1(x), F_2(y)) f_1(x) f_2(y)$, where $c$ is copula density. The copula approach provides a powerful tool for constructing a large class of multivariate distributions based on marginals from different families. Any joint distribution function may be represented through copula in which dependence structure and marginals are separately specified. For a good source on copulas, one may refer to Nelsen (2006) and Joe (2014). Copula methods could be a flexible approach for constructing a large class of bivariate lifetime distributions with the ability to cope with different kinds of data and perceive the two lifetimes for the same patient. For example, it may be of interest in the study of human organs associated with kidney or eyes, and times between the first and second hospitalization for a particular disease (see Rinne (2008); Bhattacharjee and Mishra (2016)).

In the statistical literature, many authors are used copula structure to construct a number of bivariate distributions to analyze lifetime data. Kundu and Gupta (2011) proposed an absolute continuous bivariate generalized exponential distribution via simple transformation from exchangeable distribution. The proposed distribution can be easily derived from the Clayton copula with generalized exponential distribution marginals. Several statistical properties of the proposed distribution are discussed using copula techniques. A bivariate generalized exponential distribution based on Farlie-Gumbel-Morgenstern (FGM) copula had been proposed and studied by Achcar *et al.* (2015). Recently, Kundu and Gupta (2017) proposed the bivariate Birnbaum-Saunders distribution from Gaussian copula and investigated its several reliability and dependence properties. Abd Elaal and Jarwan (2017) considered bivariate generalized exponential distributions derived from FGM and Plackett copula functions and demonstrated their applications using real data sets. A bivariate modified Weibull distribution embedded by Peres *et al.* (2018) via FGM copula. Popovic *et al.* (2018) discussed statistical properties of a bivariate Dagum distribution through copula. Nair *et al.* (2018) proposed a bivariate model for lifetime data analysis based on copula functions. Samanthi and Sepanski (2019) proposed a new bivariate extension of the beta-generated distributions using Archimedean copulas and discussed its applications in financial risk management. Shih *et al.* (2019) introduced a bivariate FGM copula model for bivariate meta-analysis and develop a maximum likelihood estimator for the common mean. De Oliveira Peres *et al.* (2020) proposed bivariate standard Weibull lifetime distributions using different copula functions and utilized them in real applications. Ota and Kimura (2021) discussed an effective algorithm for estimating the parameters of the multivariate FGM copula by using inference functions for the margins method. Several other bivariate distributions using copula have been proposed and studied in the literature. Some important references include (Saraiva *et al.* (2018); Taheri *et al.* (2018); Najarzadegan *et al.* (2019); Almetwally *et al.* (2021)). Recently, El-Morshedy et al. (2020), Almetwally and Muhammed (2020), Muhammed and Almetwally (2020), Alotaibi et al. (2021), and El-Sherpieny et al. (2021, 2022) did extensive work on different bivariate distributions. Pandey et al. (2020) presented shared inverse Gaussian frailty models for bivariate findings. Pandey et al. (2020) looked at generalized inverse Gaussian shared frailty models based on reversed hazard rates. Pandey et al. (2021) and Tyagi et al. (2021) developed distinct GL shared frailty models based on the reversed hazard rate. Pandey et al. (2020), Tyagi et al. (2021, 2022a, 2022b, 2022c), Gupta et al. (2022) and Pandey et al. (2021) developed inverse Gaussian, weighted Lindley, and GL shared frailty models, respectively.

The aim of this paper is to introduce a new bivariate inverse Topp-Leone (BITL) model and explore its various statistical properties with an application in real data. This paper is organized as follows: In Section 2, we review some basics of the univariate ITL model. With the help of the univariate ITL model, we define a new family of BITL model using the FGM copula. In Section 3, we derive the expressions for joint survival function, joint hazard rate, joint reversed hazard rate, and conditional density for the proposed BITL model.

## 2  Bivariate Inverse Topp-Leone Model

Topp and Leone (1955) illustrated the Topp-Leone (TL) model with minimal support as a conceptual model in reliability assessments. The density function of the TL model is J-shaped, whereas the hazard function is bathtub-shaped. Numerous scholars have done groundbreaking disquisition due to the relevance of the TL model. Hassan *et al.* (2020) acquired an inverse modified form of the TL model specified on the $\text{IR}^+$ domain, named the Inverted Topp-Leone model, due to the importance and relevance of inverted models with distribution function (DF), probability density function (pdf), and survival function:

$$F_X(x) = 1 - (x+1)^{-2\xi}(2x+1)^{\xi}; x \in \text{IR}^+, \xi \in \text{IR}^+, \qquad (2.1)$$

$$f_X(x) = 2\xi x(x+1)^{-(2\xi+1)}(2x+1)^{\xi-1}; x \in \text{IR}^+, \xi \in \text{IR}^+ \qquad (2.2)$$

$$\phi_X(x) = (x+1)^{-2\xi}(2x+1)^{\xi}; x \in \text{IR}^+, \xi \in \text{IR}^+, \qquad (2.3)$$

respectively. FGM copula is one of the most popular parametric families of copulas and has been widely used in literature due to its simple structure. Morgenstern (1956) proposed the FGM family and was later studied by

Gumbel (1958, 1960) using normal and exponential marginals, respectively. Farlie (1960) extended this family and derived its correlation structure, hence termed as the FGM family of distributions. The bivariate FGM copula is given by

$$C(u,v) = uv[1+\delta(1-u)(1-v)], \quad \delta \in [-1,1]. \tag{2.4}$$

In order to achieve the wider applications of the FGM copula in real applications, a large number of generalized FGM copulas have been proposed and studied in the literature. Some of the recent references include Amblard and Girard (2009) and Pathak and Vellaisamy (2016).

The bivariate distribution determined by FGM copula is

$$F(x,y) = F_1(x)F_2(y)[1+\delta(1-F_1(x))(1-F_2(y))]; \quad \delta \in [-1,1]. \tag{2.5}$$

A new family of BITL model via FGM copula is given by

$$F_{(X,Y)}(x,y) = \left(1-(x+1)^{-2\xi_1}(2x+1)^{\xi_1}\right)\left(1-(y+1)^{-2\xi_2}(2y+1)^{\xi_2}\right)$$
$$\left(1+\delta\left(\left(\frac{2(2x+1)^{\xi_1}}{(x+1)^{2\xi_1}}\right)\left(\frac{2(2y+1)^{\xi_2}}{(y+1)^{2\xi_2}}\right)\right)\right) \tag{2.6}$$

A random vector $(X,Y)$ is said to have a bivariate inverted Topp-Leone (BITL) model with parameters $\xi_1$, $\xi_2$ and $\delta$ if, its distribution function is given by (2.5), and is denoted by BITL($\xi_1, \xi_2, \delta$). This family includes a mixture of exponential and gamma distributions and may be useful in a wide class of real data.

The joint density of the BITL model $f(x,y)$ defined in (2.5) is

$$f_{(X,Y)}(x,y) = 4\xi_1\xi_2 xy(x+1)^{-(2\xi_1+1)}(2x+1)^{\xi_1-1}(y+1)^{-(2\xi_2+1)}(2y+1)^{\xi_2-1}$$
$$\left(1+\delta\left(\left(\frac{2(2x+1)^{\xi_1}}{(x+1)^{2\xi_1}}-1\right)\left(\frac{2(2y+1)^{\xi_2}}{(y+1)^{2\xi_2}}-1\right)\right)\right) \tag{2.7}$$

## 3 Reliability Properties

Statistical properties are essential in influencing whether such a bivariate distribution can be implemented to a certain type of data. The bivariate model BITL established in this study is significant because it may be used to conduct an investigation of the reliability of a system consisting of two components. As a consequence, numerous reliability functions, such as the survival function, hazard function, reversed hazard rate, and conditional distribution must be constructed. The above-mentioned reliability characteristics for the bivariate distribution are derived in ensuing subsections.

### 3.1 Survival Function

There are several ways to construct the reliability function for the bivariate distribution; we prefer to use the copula approach to express the reliability function for the BITL model by using the marginal survival function $\phi(x)$ and $\phi(y)$ where $X$ and $Y$ the random variable and selection dependence structure.

**Theorem 1.** *The joint survival function for the copula is as follows*

$$\phi(x,y) = C(\phi(x),\phi(y))$$

*where the marginal survival function $u = \phi(x)$ and $v = \phi(y)$. The reliability function of FGM-BITL based on equation (3.1)*

$$\phi(x,y) = \frac{(2x+1)^{\xi_1}(2y+1)^{\xi_2}\left(\frac{\delta\left((2x+1)^{\xi_1}(2y+1)^{\xi_2}\right)}{(x+1)^{2\xi_1}(y+1)^{2\xi_2}}+1\right)}{(x+1)^{2\xi_1}(y+1)^{2\xi_2}} \tag{3.1}$$

### 3.2 Hazard Function

**Theorem 2.** *Let $(X,Y)$ be a bivariate random vector with joint density $f(x,y)$ and survival function $\phi(x,y) = P(X \in (x,+\infty), Y > \in (y,+\infty))$. Then the bivariate hazard rate function is defined as*

$$H(x,y) = \frac{f(x,y)}{\phi(x,y)}.$$

$$H(x,y) = \frac{4\xi_1\xi_2 xy\left(\delta\left(2(x+1)^{-2\xi_1}(2x+1)^{\xi_1}-1\right)\left(2(y+1)^{-2\xi_2}(2y+1)^{\xi_2}-1\right)+1\right)}{(x+1)(2x+1)(y+1)(2y+1)\left(\delta(2x+1)^{\xi_1}(x+1)^{-2\xi_1}(y+1)^{-2\xi_2}(2y+1)^{\xi_2}+1\right)}$$

### 3.3 Reversed Hazard Rate Function

**Theorem 3.** Let $(X,Y)$ be a bivariate random vector with joint density $f(x,y)$ and distribution function $F(x,y) = P(X \in (0,x), Y \in (0,y))$. Then the bivariate reversed hazard rate function is defined as

$$m(x,y) = \frac{f(x,y)}{F(x,y)}.$$

$m(x,y) =$

$$\frac{4\xi_1\xi_2 xy(2x+1)^{\xi_1-1}(2y+1)^{\xi_2-1}\left(\delta\left((x+1)^{2\xi_1}-2(2x+1)^{\xi_1}\right)\left((y+1)^{2\xi_2}-2(2y+1)^{\xi_2}\right)+(x+1)^{2\xi_1}(y+1)^{2\xi_2}\right)}{(x+1)(y+1)\left((x+1)^{2\xi_1}-(2x+1)^{\xi_1}\right)\left((y+1)^{2\xi_2}-(2y+1)^{\xi_2}\right)\left(\delta(2x+1)^{\xi_1}(2y+1)^{\xi_2}+(x+1)^{2\xi_1}(y+1)^{2\xi_2}\right)}$$

### 3.4 Conditional Distribution

**Theorem 4.** Let $(X,Y) \sim \text{BITL}(\xi_1, \xi_2, \delta)$. Then

i.     The $X \sim \text{ITL}(\xi_1)$. and $Y \sim \text{ITL}(\xi_2)$

ii.     the conditional density of $X$ given $Y = y$ is

$$f(x|y) = 2\xi_1 x(x+1)^{-2\xi_1-1}(2x+1)^{\xi_1-1}$$
$$\left(\delta\left(2(x+1)^{-2\xi_1}(2x+1)^{\xi_1}-1\right)\left(2(y+1)^{-2\xi_2}(2y+1)^{\xi_2}-1\right)+1\right)$$

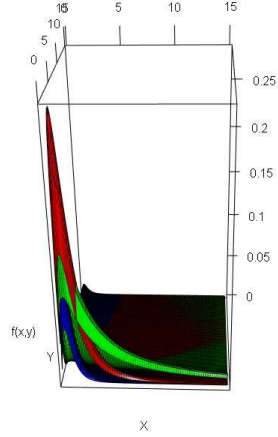

Figure 1: PDF BITL model

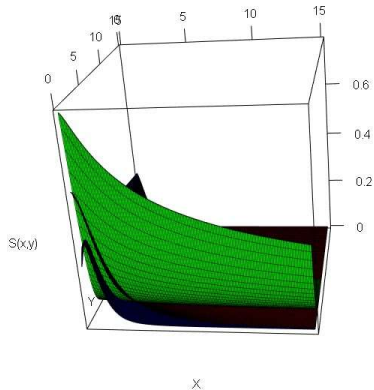 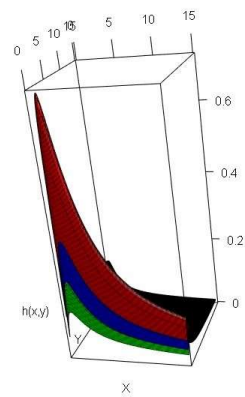

Figure 2: Survival BITL model    Figure 3: Hazard BITL model

**References**

Abd Elaal MK, Jarwan RS. Inference of bivariate generalized exponential distribution based on copula functions. Applied Mathematical Sciences. (2017); 11(24): 1155-1186.

Achcar JA, Moala FA, Tarumoto MH, Coladello LF. A bivariate generalized exponential distribution derived from copula functions in the presence of censored data and covariates. Pesquisa Operacional. (2015); 35: 165-186.

Alley W. The Palmer drought severity index: limitations and assumptions. J Clim Appl Meteorol. (1984); 23:1100-1109.

Almetwally EM, Sabry MA, Alharbi R, Alnagar D, Mubarak SA, Hafez, EH. Marshall–Olkin Alpha Power Weibull Distribution: Different Methods of Estimation Based on Type-I and Type-II Censoring. Complexity. (2021).

Almetwally EM, Muhammed, HZ. On a bivariate Fréchet distribution. J Stat Appl Probab. (2020); 9(1): 1-21.

Alotaibi R, Khalifa M, Almetwally EM, Ghosh I. Classical and Bayesian Inference of a Mixture of Bivariate Exponentiated Exponential Model. Journal of Mathematics. (2021). https://doi.org/10.1155/2021/5200979

Amblard C, Girard S. A new extension of bivariate FGM copulas. Metrika. (2009); 70(1): 1-17.

Anderson JE, Louis TA, Holm NV, Harvald B. Time-dependent association measures for bivariate survival distributions. Journal of the American Statistical Association. (1992); 87(419): 641-650.

Bhattacharjee S, Misra SK. Some aging properties of Weibull models. Electronic Journal of Applied Statistical Analysis. (2016); 9(2): 297-307.

Dasgupta, R. On the distribution of Burr with applications. Sankhya B. (2011): 73: 1-19.

de Oliveira Peres MV, Achcar JA, Martinez EZ. Bivariate lifetime models in presence of cure fraction: a comparative study with many different copula functions. Heliyon. (2020); 6(6): e03961.

Dolati A, Amini M, Mirhosseini SM. Dependence properties of bivariate distributions with proportional (reversed) hazards marginals. Metrika. (2014); 77(3): 333-347.

Eberly LE, Casella G. Estimating Bayesian credible intervals. Journal of Statistical Planning and Inference. (2003); 112 (1–2):115–32. doi:10.1016/S0378-3758(02)00327-0.

El-Morshedy M, Alhussain ZA, Atta D, Almetwally EM, Eliwa MS. Bivariate Burr X generator of distributions: properties and estimation methods with applications to complete and type-II censored samples. Mathematics. (2020); 8(2): 264.

El-Sherpieny ESA, Almetwally EM, Muhammed HZ. Bayesian and non-bayesian estimation for the parameter of bivariate generalized Rayleigh distribution based on clayton copula under progressive type-II censoring with random removal. Sankhya A. (2021); 1-38. https://doi.org/10.1007/s13171-021-00254-3

El-Sherpieny ESA, Muhammed HZ, Almetwally EM. Progressive Type-II Censored Samples for Bivariate Weibull Distribution with Economic and Medical Applications. Annals of Data Science. (2022); 1-35. https://doi.org/10.1007/s40745-022-00375-y

Farlie DJ. The performance of some correlation coefficients for a general bivariate distribution. Biometrika. (1960); 47(3/4): 307-323.

Gumbel EJ. Statistics of extremes. Columbia university press. (1958).

Gumbel EJ. Bivariate exponential distributions. Journal of the American Statistical Association. (1960); 55(292): 698-707.

Gupta, P., Pandey, A., Tyagi, S.. Comparison of Multiplicative Frailty Models under Generalized Log-Logistic-II Baseline Distribution for Counter Heterogeneous Left Censored Data, (2022), 1, 97-114.

Hassan AS, Elgarhy M, Ragab R. Statistical properties and estimation of inverted Topp-Leone distribution. J. Stat. Appl. Probab. (2020); 9(2): 319-331.

Ibrahim JG, Ming-Hui C, Sinha, D. Bayesian Survival Analysis. Springer, Verlag. (2001).

Joe H. Multivariate models and multivariate dependence concepts. Chapman and Hall, New York. (1997).

Joe H. Dependence modeling with copulas. CRC press. (2014).

Johnson NL, Kotz SI, Balakrishnan N. Beta distributions. Continuous univariate distributions. 2nd ed. New York, NY: John Wiley and Sons. (1994); 221-235.

Johnson NL, Kotz S. A vector multivariate hazard rate. Journal of Multivariate Analysis. (1975); 5(1): 53-66.

Kundu D, Gupta AK. On bivariate inverse Weibull distribution. Brazilian Journal of Probability and Statistics. (2017); 31(2): 275-302.

Kundu D, Gupta, RC. On Bivariate Birnbaum–Saunders Distribution. American Journal of Mathematical and Management Sciences. (2017); 36(1): 21-33.

Kundu D, Gupta, RD. Bivariate generalized exponential distribution. Journal of multivariate analysis 100, no. 4 (2009): 581-593.

Kundu D, Gupta, RD. Absolute continuous bivariate generalized exponential distribution. AStA Advances in Statistical Analysis. (2011); 95(2): 169-185.

Marshall AW, Olkin I. A generalized bivariate exponential distribution. Journal of applied probability. (1967); 4(2): 291-302.

Mirhosseini SM, Amini M, Kundu D, Dolati A. On a new absolutely continuous bivariate generalized exponential distribution. Statistical Methods & Applications. (2015); 24(1): 6183.

Morgenstern D. Einfache beispiele zweidimensionaler verteilungen. Mitteilingsblatt fur Mathematische Statistik. (1956); 8: 234-235.

Muhammed HZ, Almetwally EM. Bayesian and non-Bayesian estimation for the bivariate inverse weibull distribution under progressive type-II censoring. Annals of Data Science. (2020); 1-32. https://doi.org/10.1007/s40745-020-00316-7

Nair NU, Sankaran PG, John P. Modelling bivariate lifetime data using copula. Metron. (2018); 76(2): 133-153.

Najarzadegan H, Alamatsaz MH, Kazemi I. Discrete Bivariate Distributions Generated By Copulas: DBEEW Distribution. Journal of Statistical Theory and Practice. (2019); 13(3): 1-30.

Nadarajah S. A bivariate pareto model for drought. Stochastic Environmental Research and Risk Assessment. (2009); 23(6): 811-822.

Nelsen RB. Springer series in statistics, An introduction to copulas. (2006).

Norstrom JG. The use of precautionary loss functions in risk analysis. IEEE Transactions on reliability. (1996); 45(3): 400-403.

Oakes D. Bivariate survival models induced by frailties. Journal of the American Statistical Association. (1989); 84(406): 487-493.

Ota S, Kimura M. Effective estimation algorithm for parameters of multivariate Farlie–Gumbel–Morgenstern copula. Japanese Journal of Statistics and Data Science. (2021); 1-30.

Pandey, A., Hanagal, D. D., Gupta, P., and Tyagi, S.. Analysis of Australian Twin Data Using Generalized Inverse Gaussian Shared Frailty Models Based on Reversed Hazard Rate. International Journal of Statistics and Reliability Engineering, (2020); 7(2), 219-235.

Pandey, A., Bhushan, S., Pawimawha, L. and Tyagi, S.. Analysis of Bivariate Survival Data using Shared Inverse Gaussian Frailty Models: A Bayesian Approach, Predictive Analytics Using Statistics and Big Data: Concepts and Modeling, Bentham Books, (2020); (14), 75-88.

Pandey, A., Hanagal, D. D., Tyagi, S., and Gupta, P.. Generalized Lindley Shared Frailty Based on Reversed Hazard Rate. International Journal of Reliability, Quality and Safety Engineering, (2021); 2150040.

Pandey, A., Hanagal, D. D., and Tyagi, S.. Generalized Lindley Shared Frailty Models. Statistics and Applications, (2021); 19(2), 41-62.

Pandey, A., Tyagi, S. Comparison of Multiplicative Frailty Models Under Weibull Baseline Distribution. Lobachevskii J Math 42, 3184–3195 (2021). https://doi.org/10.1134/S1995080222010140

- Pathak AK, Vellaisamy P. Various measures of dependence of a new asymmetric generalized Farlie–Gumbel–Morgenstern copulas. Communications in Statistics-Theory and Methods. (2016); 45(18): 5299-5317.
- Peres MVDO, Achcar JA, Martinez EZ. Bivariate modified Weibull distribution derived from Farlie-Gumbel-Morgenstern copula: a simulation study. Electronic Journal of Applied Statistical Analysis. (2018); 11(2): 463-488.
- Popović BV, Genc AI, Domma F. Copula-based properties of the bivariate Dagum distribution. Computational and Applied Mathematics. (2018); 37(5): 6230-6251.
- Rinne H. The Weibull distribution: a handbook. CRC press. (2008)
- Samanthi RG, Sepanski J. A bivariate extension of the beta generated distribution derived from copulas. Communications in Statistics-Theory and Methods. (2019); 48(5): 1043-1059.
- Sankaran PG, Nair NU. A bivariate Pareto model and its applications to reliability. Naval Research Logistics (NRL). (1993); 40(7): 1013-1020.
- Santos CA, Achcar JA. A Bayesian analysis for multivariate survival data in the presence of covariates. Journal of Statistical Theory and Applications. (2010); 9: 233-253.
- Saraiva EF, Suzuki AK, Milan, LA. Bayesian computational methods for sampling from the posterior distribution of a bivariate survival model, based on AMH copula in the presence of right-censored data. Entropy. (2018); 20(9): 642.
- Sarhan AM, Hamilton DC, Smith B, Kundu D. The bivariate generalized linear failure rate distribution and its multivariate extension. Computational statistics & data analysis. (2011); 55(1): 644-654.
- Sarhan AM, Kundu D. Generalized linear failure rate distribution. Communications in Statistics-Theory and Methods. (2009); 38(5): 642-660.
- Sen S, Chandra N, Maiti SS. On properties and applications of a two-parameter XGamma distribution. Journal of Statistical Theory and Applications. (2018); 17(4): 674-685.
- Shih JH, Konno Y, Chang YT, Emura T. Estimation of a common mean vector in bivariate meta-analysis under the FGM copula. Statistics. (2019); 53(3): 673-695.
- Sklar M. Fonctions de repartition an dimensions et leurs marges. Publ. inst. statist. univ. Paris. (1959) 8: 229-231.
- Taheri B, Jabbari H, Amini M. Parameter estimation of bivariate distributions in presence of outliers: An application to FGM copula. Journal of Computational and Applied Mathematics. (2018); 343: 155-173.
- Topp CW, Leone FC. A family of J-shaped frequency functions. Journal of the American Statistical Association. (1955); 50(269): 209-219.
- Yevjevich V. An objective approach to definitions and investigations of continental hydrologic droughts. Hydrologic paper no. 23. Colorado State University, Fort Collins. (1967).

Tyagi, S., Pandey, A., Hanagal, D. D., and Gupta, P.. Bayesian inferences in generalized Lindley shared frailty model with left censored bivariate data. Advance Research Trends in Statistics and Data Science, (2021); 137–157.

Tyagi, S., Pandey, A., Agiwal, V., and Chesneau, C.. Weighted Lindley multiplicative regression frailty models under random censored data. Computational and Applied Mathematics, (2021); 40(8), 1-24.

Tyagi, S., Pandey, A. & Chesneau, C. Identifying the Effects of Observed and Unobserved Risk Factors Using Weighted Lindley Shared Regression Model. J Stat Theory Pract 16, 16 (2022). https://doi.org/10.1007/s42519-021-00241-9

Tyagi, S., Pandey, A. & Chesneau, C. Weighted Lindley Shared Regression Model for Bivariate Left Censored Data. Sankhya B (2022). https://doi.org/10.1007/s13571-022-00278-1

Tyagi, S., Pandey, A., and Hanagal, D.D.. "Shared Frailty Models Based on Cancer Data." arXiv preprint arXiv:2112.10986 (2021).